# Observation of hybrid higher-order skin-topological effect in non-Hermitian topolectrical circuits

Deyuan Zou[1*], Tian Chen[1*], Wenjing He[2], Jiacheng Bao[2], Ching Hua Lee[3], Houjun Sun[2$], and Xiangdong Zhang[1+]

[1]Key Laboratory of advanced optoelectronic quantum architecture and measurements of Ministry of Education, Beijing Key Laboratory of Nanophotonics & Ultrafine Optoelectronic Systems, School of Physics, Beijing Institute of Technology, 100081, Beijing, China

[2] Beijing Key Laboratory of Millimeter wave and Terahertz Techniques, School of Information and Electronics, Beijing Institute of Technology, Beijing 100081, China

[3]Department of Physics, National University of Singapore, Singapore 117542

*These authors contributed equally to this work. [+$]Author to whom any correspondence should be addressed. E-mail: zhangxd@bit.edu.cn; sunhoujun@bit.edu.cn; chentian@bit.edu.cn

**Abstract**

Robust boundary states epitomize how deep physics can give rise to concrete experimental signatures with technological promise. Of late, much attention has focused on two distinct mechanisms for boundary robustness - topological protection, as well as the non-Hermitian skin effect. In this work, we report the first experimental realizations of hybrid higher-order skin-topological effect, in which the skin effect selectively acts only on the topological boundary modes, not the bulk modes. Our experiments, which are performed on specially designed non-reciprocal 2D and 3D topolectrical circuit lattices, showcases how non-reciprocal pumping and topological localization dynamically interplays to form various novel states like 2D skin-topological, 3D skin-topological-topological hybrid states, as well as 2D and 3D higher-order non-Hermitian skin states. Realized through our highly versatile and scalable circuit platform, theses states have no Hermitian nor lower-dimensional analog, and pave the way for new applications in topological switching and sensing through the simultaneous non-trivial interplay of skin and topological boundary localizations.

Much of contemporary condensed matter physics has been dominated by theoretical and experimental investigations into robust boundary phenomena. Originally formulated in the context of anomalies, they revolutionized the field of condensed matter physics in the form of topological insulators and semimetals [1-3]. More recently, they aroused much attention again as non-Hermitian skin states, which demonstrated how unbalanced non-Hermitian gain/loss can challenge well-held tenets of bulk-boundary correspondence [4-12].

While skin and topological effects are already by themselves conceptually deep, with contrasting implications on the bulk-boundary correspondence, their simultaneous nontrivial interplay has been particularly intriguing. Introduced in Ref.[13], hybrid skin-topological states represent a hierarchy of novel higher-dimensional new states without analogs in Hermitian or non-topological settings. Characterized by scenarios where topological localization dynamically allows the non-Hermitian skin effect to act only on certain topological modes, they greatly augment the richness of higher-dimensional robustness beyond higher-order topological phenomena, and provokes the re-formulation of higher-order topological bulk-boundary correspondence to accommodate various avenues of higher-order skin and topological interplay [14, 15]. Their non-triviality and richness is also apparent in the context of applications: this simultaneous interplay of the skin effect and higher-order topology has been proposed to be used as a topological switch [16].

In these advances, theory has always preceded experiments. The reason is because most tight-binding models proposed are rather artificial for realization in conventional materials or metamaterials. While higher-order topological phenomena [17-23] and the non-Hermitian skin effect [9-12] have been separately realized in several high-profile experiments, their combined interplayed has so far remained a theoretical fantasy due to concomitant challenges of high dimensionality, artificial sublattice structure and non-Hermitian instabilities.

In this work, we demonstrate the first experimental realization of hybrid higher order skin-topological states in 2D and 3D through a topolectrical circuit platform. Electrical circuits are ideally suited for transcending the abovementioned challenges, since electronic components, which have benefitted from industrial refinement over the decades, can accommodate almost any desired features like arbitrary long-range connectivity, dimensionality, non-Hermiticity gain/loss as well as non-reciprocity. Recently, simulating topological states with electric circuits has attracted lots of interests based on the similarity between circuit Laplacian and lattice Hamiltonian [23-30]. Some topological states have been observed in circuit networks [31-35]. Through a combination of z-direction non-Hermitian INICs and 2D topological circuits, here we managed to achieve a 3D realization of not just the non-Hermitian skin effect, but also a network of competing such effects that conspire to result in higher-order hybrid states, much beyond the scope of previous circuit demonstrations of 1D skin or higher-order topological robustness individually [10, 18].

**Hybrid higher-order topological skin effect in 2D topolectrical circuits**

Hybrid skin-topological phenomena represent not just the simultaneous presence of non-Hermitian skin as well as topological localizations. They are special scenarios where topological localization in one direction dynamically "switches on" the skin effect in another direction, thereby allowing the skin effect to be felt only by topological modes. Given the wide variety of possible types of topological modes, as well as various nontrivial ways whereby the skin effect modifies topological properties, even in 1 dimension, hybrid modes thus comes in a vast array of possibilities. This is especially interesting in 2 dimensions or higher, which we also experimentally probe, because even topological localization per se is subject to higher-order topological effects, which can be modified by skin localization even before they interplay as hybrid phenomena.

We firstly provide the theoretical design of 2D electric circuits to observe skin-topological (ST) and skin-skin (SS) modes, and then give the experimental results to demonstrate such a design. Voltage measurements are particularly sensitive to the spectrum of the circuit Laplacian. In matrix form, Kirchhoff's law is expressed as $I = JV$, where $J$ is the circuit Laplacian. The electrical potentials $V$ at each node can be obtained by inverting this expression to obtain

$$V = J^{-1}I = \sum_\mu \varepsilon_\mu^{-1} |\psi_\mu\rangle \langle\psi_\mu| I \qquad (1)$$

Notably, the potentials $V$ are most sensitive to small eigenvalues $\varepsilon_\mu$, especially if they are vanishing. In the case of skin-topological hybrid modes, the hybrid mode with zero eigenvalue thus dominates voltage measurements, leading the voltage profile $V$ proportional to the corresponding hybrid eigenmode. Indeed, we experimentally observe that the voltage profile is topologically localized in the y-direction, and skin-localized in the x-direction, even though the bulk modes are not supposed to even experience the skin effect.

The designed 2D electric circuit network is shown in Fig.1(a). The sample contains 6×6

units and each unit cell contains four sublattices (a, b, c, d). In the network, different circuit unit cells can be used to construct systems with different functions. For example, if we use the unit cell as shown in Fig. 1(b), the hybrid second-order ST modes appear. In contrast, if the unit cell as shown in Fig. 1(c) is used, the SS modes can be observed. Two kinds of unit cell are composed of capacitances, inductances and three different kinds of negative impedance converter through current inversion (INIC). Operational amplifiers arranged as INIC allow the type of nonreciprocity in the circuit to be precisely tuned, where their detail structures are shown in Fig.1(d).

For the unit cell as shown in Fig.1(b), we make the directions of INIC be opposite along x and y directions. This can result in vanished net nonreciprocity since the nonreciprocities cancel along x and y directions, but local nonreciprocity among four sublattices in each unit cell still exists, which corresponds to the 2D hybrid lattice model revealed in Ref. [13]. If the directions of INIC are the same along x and y directions as shown in Fig.1(c), the nonreciprocities along both directions of the electric circuit do not interfere destructively, which accomplishes 2D skin lattice model. Details of the theoretical analysis for the lattice models are provided in Methods. No matter what kind of circuit, we can derive circuit Laplacian $J_{2D}(\omega)$ in the moment space at the resonance frequency based on Kirchhoff's current law. It can be written as

$$J_{2D}(\omega) = i\omega \begin{bmatrix} 0 & 0 & C_1 - C_3 + Ce^{-iq_x} & C_1 \pm C_2 + Ce^{-iq_y} \\ 0 & 0 & -C_1 - C_2 - Ce^{iq_y} & C_1 \pm C_3 + Ce^{iq_x} \\ C_1 + C_3 + Ce^{iq_x} & -C_1 + C_2 - Ce^{-iq_y} & 0 & 0 \\ C_1 \mp C_2 + Ce^{iq_y} & C_1 \mp C_3 + Ce^{-iq_x} & 0 & 0 \end{bmatrix}, \quad (2)$$

where $C = 2.2\text{nF}$, $C_1 = 1\text{nF}$, $C_2 = 820\text{pF}$ and $C_3 = 390\text{pF}$. If we choose – in $\pm$ and + in $\mp$ from the matrix above, Eq.(2) represents the circuit Laplacian for the case where the hybrid second-order ST effect can be shown. Otherwise, it corresponds to the case that can show the SS effect. The detailed derivation of Eq. (2) is given in Methods.

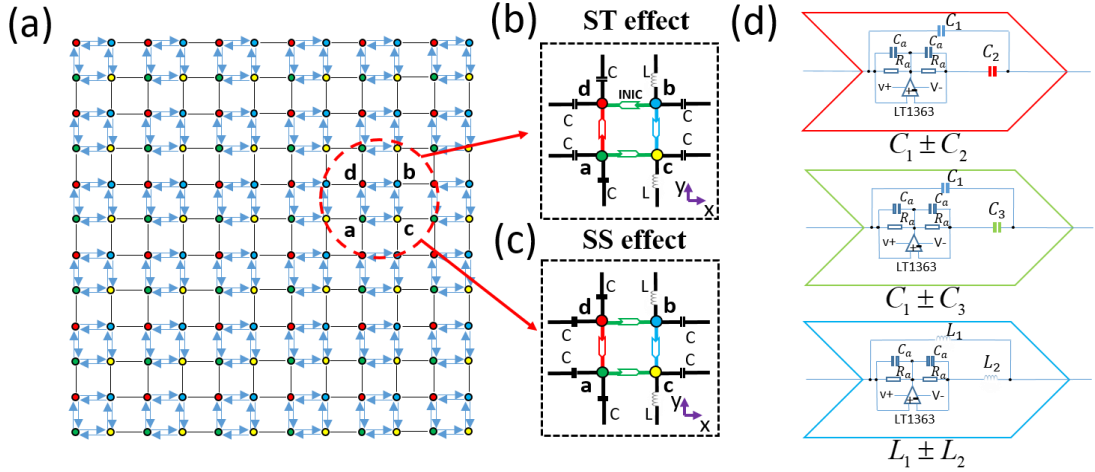

**Fig. 1. The circuit design for the hybrid second-order ST and SS effects.** (a) Designed 2D electric circuit network with different values of intra-cell and inter-cell couplings. Each unit cell contains four sublattices (a, b, c, d). The blue lines with arrows represent INICs which can tune nonreciprocity in the circuit and the black lines denote inter-cell capacitive and inductive couplings. (b) and (c) display the circuit unit for the realization of hybrid second-order ST effect and SS effect with $C = 2.2\text{nF}$ and $L = 1.5\text{uH}$, respectively. Different colors of INIC indicate different values of non-Hermitian coupling. (d) Detail structures of different INICs with $C_1 = 1\text{nF}$, $L_1 = 3.3\text{uH}$, $R_a = 20\Omega$ and $C_a = 1\text{uF}$.

As a concrete illustration, and to foreshadow our experimental results, we firstly display the spectrum of our resonant circuit Laplacian for the 2D ST case in Fig.2(a). While topological localization only causes the appearance of special, non-extensively scaling boundary modes when a boundary appears, the non-Hermitian skin effect is characterized by a modification of the entire spectrum upon opening up a boundary. In the ST case, a small number of topological modes, arranged in the form of a spectral loop (blue), appears when y-boundaries are implemented. When x-boundaries are further introduced, we manifest observe that only these topological modes acquire very different spectral values (green) i.e. undergo the skin effect; the other bulk modes are unchanged by the introduction of either boundary. In particular, there is a ST zero-mode that arises only when the skin effect selectively applies to the topological modes in this manner. This zero mode will contribute massively to the experimentally observed electrical potentials, as further elaborated. By contrast, in Fig.2(b) the entire spectrum is dramatically altered whether the x-boundary or y-boundary is introduced, indicative of the skin effect in both directions.

We next perform numerical simulations using LTspice. In the calculation, we take $C=2.2\text{nF}$, $C_1=1\text{nF}$, $C_2=820\text{pF}$, $C_3=390\text{pF}$, $L=1.5\text{uH}$, $L_1=3.3\text{uH}$, $L_2=3.9\text{uH}$ and $L_3=8.2\text{uH}$. By appropriate grounding design, the circuit has a same resonance angular frequency $\omega_0=1/\sqrt{LC}=1/\sqrt{L_1C_1}=1/\sqrt{L_2C_2}=1/\sqrt{L_3C_3}$. We excite the two circuits in same positions which are red stars in Fig.2 (c) and (d). Actually, points in line 11 are all suitable for exciting. The resulting voltage distribution at the resonance frequency $f_0=2.77$ MHz on this electric circuit is presented in Fig. 2(c) and(d) for two cases, respectively. It can be seen clearly in Fig. 2(c) that the large amplitude of voltages can be found at the top left corner and lower right corner. Moreover, the amplitudes on sublattices in unit cell plaquette are unequal. For example, at the top left corner of circuit, only the sublattice d of plaquettes possess large amplitude, and the neighboring sublattices a and b of plaquettes have very small amplitudes. For comparison, at the lower right corner of circuit, only the sublattice c of plaquettes displays large amplitude, and the neighboring sublattices a and b of plaquettes also have very small amplitudes. These unequal distributions at different sublattices indicate that the corner modes stem from the nonreciprocal skin effect on the 1D topological modes along x and y directions. They are locally nonreciprocal from the lack of full destructive interference. Such spontaneous breaking of sublattice symmetry and hence nonreciprocity is generic among topological modes, which gives rise to hybrid ST modes shown in Ref. [13]. In Fig. 2(d), the corner modes are indicated by large-amplitude voltages at the lower right corner in the circuit. The voltages at different sublattices of the lower right corner are nearly the same. This is typical second order skin effect which has the similar feature with the 1D skin mode shown in Ref. [10].

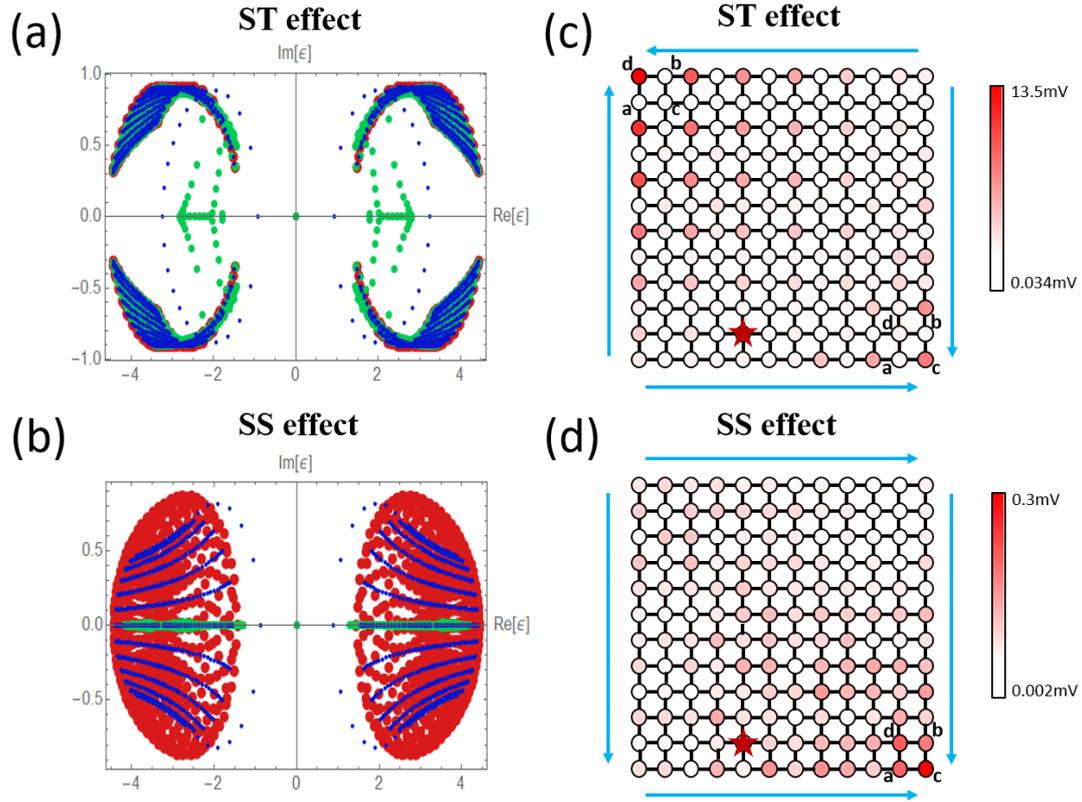

**Fig.2. Simulated results for the hybrid second-order ST and SS effects.** (a)-(b) Spectrum of our resonant circuit Laplacian for the ST and SS effects. Red, blue and green dots represent the spectrum of the circuit Laplacian subject to double periodic boundary conditions, x-periodic, y-open boundary conditions, and double open boundary conditions respectively. For hybrid ST effect in (a), when going from double periodic boundaries to y-open boundaries (red to blue), only $1/Ly$ of the eigenvalues change drastically, signature of topological boundary modes. When next going from y-open boundaries to double open boundaries (blue to green), only topological eigenvalues in the blue loop morph into the loop interior (undergo the skin effect), but not the majority bulk modes. For SS effect in (b), the spectra in all these three boundary conditions differ completely, signify skin effects from both x and y boundaries. (c) and (d) show the voltage distributions at the resonance frequency for hybrid second-order ST effect and SS effect, respectively. Big arrows in (c) show the directions of skin and topological pumping. Big arrows in (d) show the directions of skin pumping. Red stars are the points of voltage excitation.

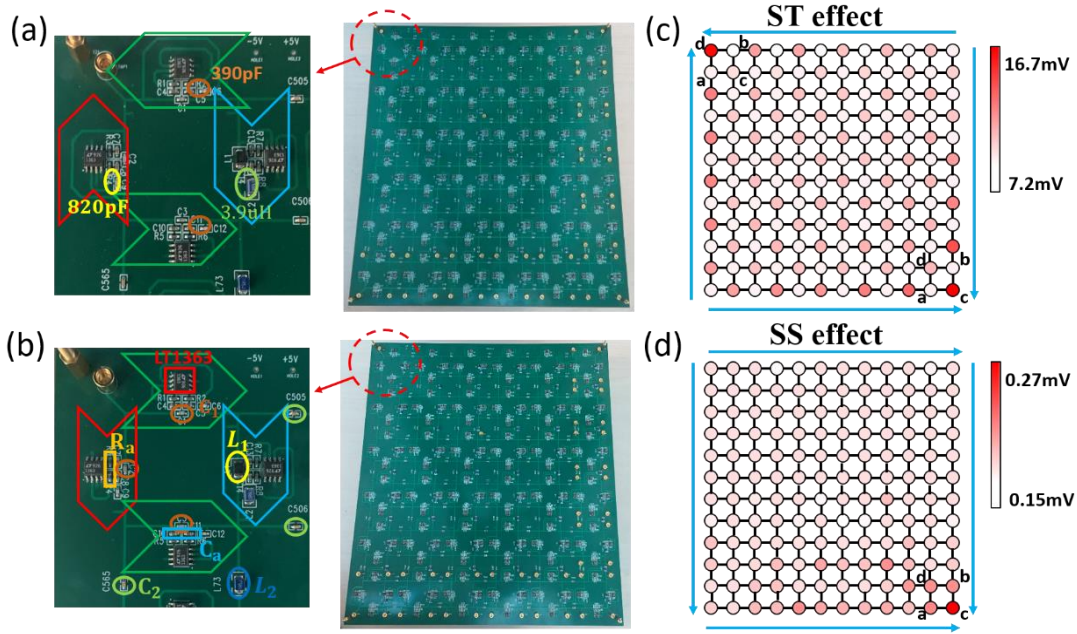

**Fig. 3. The photograph of fabricated sample and experimental results for the hybrid second-order ST effect and SS effect.** (a)-(b) The photograph of the fabricated electric circuit for the hybrid second-order ST effect(a) and SS effect(b). The inset presents the enlarged view of the unit cell. (c)-(d) The measured voltage distribution at the resonance frequency for the hybrid second-order ST effect and SS effect in (c) and (d), respectively. Big arrows in (c) show the directions of skin and topological pumping. Big arrows in (d) show the directions of skin pumping.

To experimentally test the theoretical analysis, we fabricate two kinds of electric circuits as shown in Fig.3(a) and 3(b). The insets in Fig. 3(a) and 3(b) display the circuit structure where INICs and LC elements to achieve intra- and inter-cell couplings are marked, which correspond exactly to the unit cells presented in the above theory. Thus, the circuits in Fig. 3(a) and 3(b) also correspond exactly to the designed structures in Fig.1. The sample contains 6×6 units and the parameters are the same to the above theoretical designs. It is worthy to note that the tolerance of the circuit elements is only 1% to avoid the detuning of corner resonance. To ensure the effective excitation of the circuit, NI PXIe-8840 Quad-Core Embedded Controller is used to tune the excitation amplitude and frequency. Details of the sample fabrication and experimental measurements are provided in Methods.

The measured voltage distributions at the resonance frequency are shown in Fig. 3(c) and 3(d). It is noted that, the dominant voltage signals on sublattices in the unit cell plaquette are similar to the theoretical results shown in Fig.2(c) and 2(d). The detailed comparisons for theoretical and experimental results are provided in S1 of Supplementary Materials. It is found that the experimental phenomena correspond exactly to the theoretical results, which indicates that the ST and SS modes have been observed successfully in designed circuit systems.

**3D hybrid skin-topological modes in topolectrical circuits**

The above discussions only focus on the 2D cases. In fact, more plentiful hybridizations of topology and nonreciprocity can appear in higher dimensions. In the following, we illustrate two kinds of corner modes in 3D electric circuits. One is caused by the z-directional nonreciprocal pumping (can show hybrid 3D skin-topological-topological (STT) effect), and the other is obtained from the nonreciprocities along x, y and z three directions (can show 3D skin-skin-skin (SSS) effect).

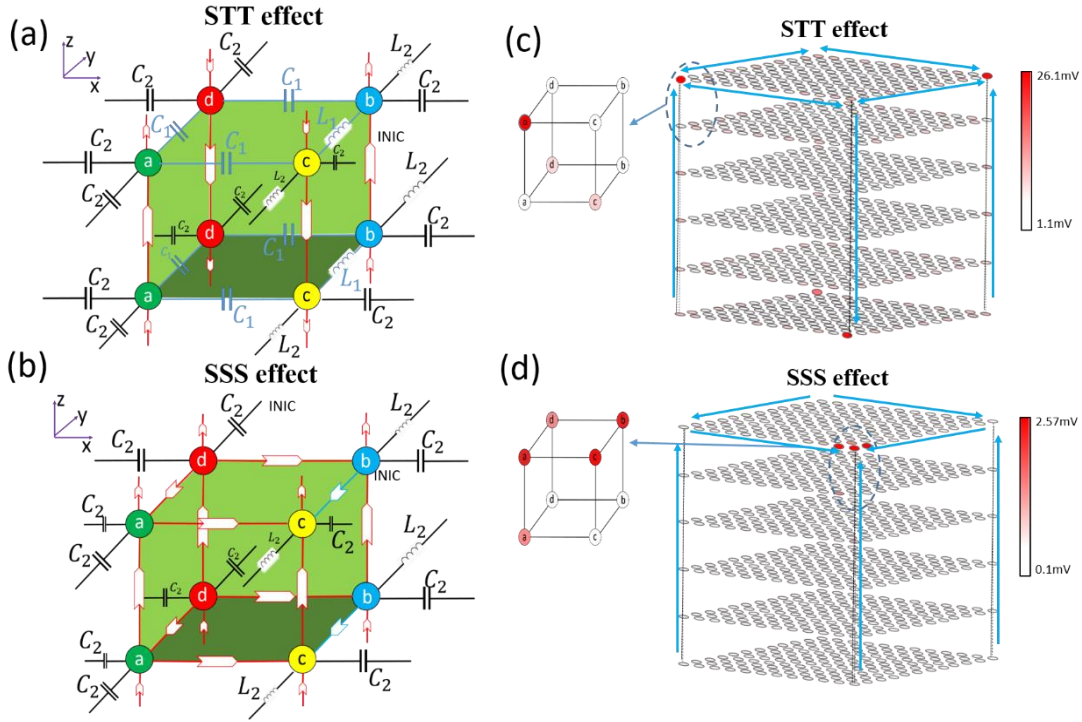

**Fig. 4. The circuit design and results for the hybrid 3D STT effect and 3D SSS corner effect.** (a)-(b) The schematic diagrams for the unit cell of the total circuit with $C_1 = 1\text{nF}, L_1 = 3.3\text{uH}, C_2 = 2.2\text{nF}$ and $L_2 = 1.5\text{uH}$. (a) is for hybrid 3D STT effect and (b) is for 3D SSS effect. The details of INICs are in Fig.1(d). (c)-(d) The voltage distributions at the resonance frequency for hybrid 3D STT effect and 3D SSS effect in (c) and (d), respectively. Big arrows in (c) show the directions of skin and topological

pumping. Big arrows in (d) show the directions of skin pumping.

The corresponding 3D circuit unit cells for two kinds of case are shown in Fig. 4(a) and 4(b), respectively. The unit cells consist of capacitances, inductances and two different kinds of INICs. In the first case (Case I) shown in Fig.4(a), the 2D layer on the x-y surface corresponds to the 2D topological lattice. To yield the nonreciprocity, the INICs are put along the positive direction of z axis in sublattices a and b of plaquettes, and reversed along the z axis in sublattices c and d. For the second case (Case II) shown in Fig.4(b), the 2D layer on the x-y surface is the same to that in Fig. 1(c). The INICs are all put forward along the positive direction of z axis which also yield the nonreciprocity. For the two kinds of case above, we can derive circuit Laplacian $J_{3D}(\omega)$ in the moment space at the resonance frequency based on Kirchhoff's current law. It can be written as

$$J_{STT3D(SSS3D)}(\omega) = J_{TT2D(SS2D)}(\omega)$$
$$+i\omega \begin{bmatrix} (C_1-C_2)e^{iq_z}+(C_1+C_2)e^{-iq_z} & 0 & 0 & 0 \\ 0 & (C_1-C_2)e^{iq_z}+(C_1+C_2)e^{-iq_z} & 0 & 0 \\ 0 & 0 & (C_1 \pm C_2)e^{iq_z}+(C_1 \mp C_2)e^{-iq_z} & 0 \\ 0 & 0 & 0 & (C_1 \pm C_2)e^{iq_z}+(C_1 \mp C_2)e^{-iq_z} \end{bmatrix}$$

(3)

If we choose + in $\pm$ and − in $\mp$ from the matrix above, Eq.(3) represents the circuit Laplacian for Case I. Otherwise, it corresponds to Case II. The $J_{TT2D(SS2D)}$ represents circuit Laplacian on x-y surface for Case I (Case II), and the remaining term in $J_{STT3D(SSS3D)}$ represents the coupling between different layers along the z direction. The detailed derivation can be found in Methods.

Similar to the 2D cases, we can simulate the circuit using LTspice. The results are shown in Fig.4(c) and 4(d) for Case I and Case II, respectively. The parameters of components are the same to those of the 2D sample. In Fig.4(c), at the top layer, the large-amplitude voltages appear only at the sublattice a of lower left corner and at the sublattice b of top right corner. At the bottom layer, the large-amplitude voltages appear only at the sublattice d of top left corner and at the sublattice c of lower right corner. The unequal distributions at different sublattices indicate the topological origin of these corner modes, which are realized by stacking 2D layers of 2nd-order Hermitian topological mode. The INICs along the z direction lead to the nonreciprocal skin mode. So these corner modes at top and bottom layers come from the interplay between the topological and nonreciprocal pumping, which gives rise to STT modes described in Ref. [13]. In Fig.4(d), the large-amplitude voltages emerge at the lower right corner of the top layer. Moreover, nearly the same amplitudes of voltages on different sublattices at the lower right corner of the top layer indicate the nonreciprocal origin of these corner modes. In such settings, nonreciprocal pumping leads to accumulation of boundary skin modes along each direction, that is, the appearance of 3D SSS modes. So no matter in 2D or 3D, the corner modes (2D ST and 3D STT modes) induced by both topological and nonreciprocal pumping lead to unequal distributions of voltages on sublattices; for comparison, the corner modes caused by only nonreciprocal pumping bring about the nearly same distributions of voltages on sublattices.

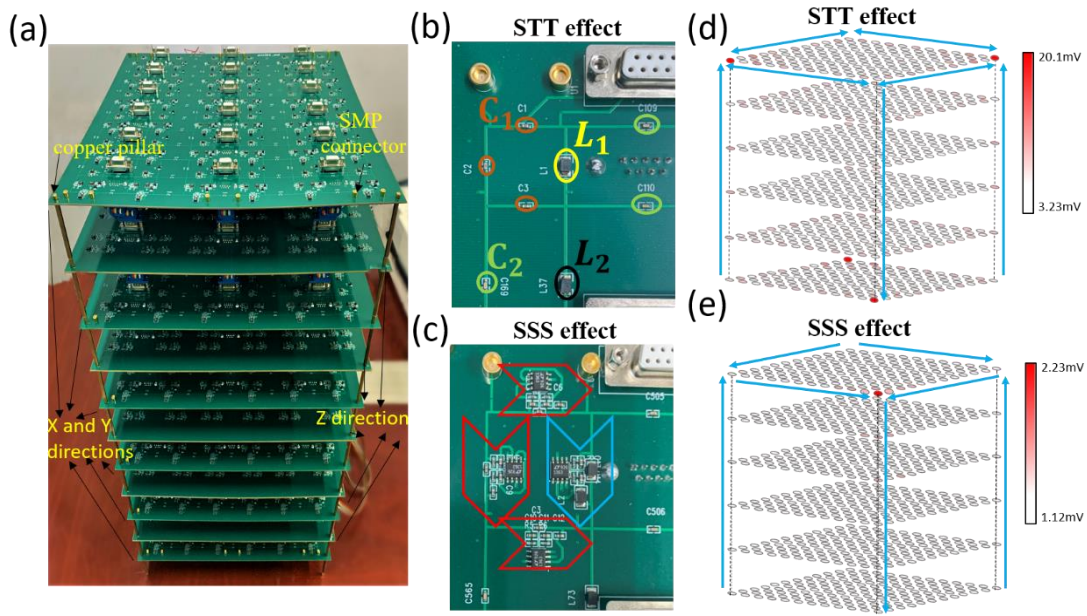

**Fig. 5. The photograph of fabricated sample and experimental results for the hybrid 3D STT effect and 3D SSS effect.** (a) The photograph of the electric circuit of the 3D SSS corner effect. (b)-(c) The enlarged view of the unit cells for the hybrid 3D STT effect in (b) and the 3D SSS effect in (c). (d)-(e) The voltage distributions at the resonance frequency for hybrid 3D STT effect and 3D SSS effect in (d) and (e), respectively. Big arrows in (c) show the directions of skin and topological pumping. Big arrows in (d) show the directions of skin pumping.

To experimentally observe Case I and Case II, we fabricate two kinds of electric circuits. The photograph image of the fabricated sample is shown in Fig. 5(a). Due to the size limit of the PCB fabrication in the 3D sample, we cut the whole sample into eleven pieces. Six of them are used to describe the interaction on x-y surface of circuit designs, and the remaining five pieces contain the couplings in the circuit designs along the z direction. Details of the z-direction connections are given in S3 of Supplementary Materials. The enlarged views of the unit cell are shown in Fig. 5(b) and 5(c) for Case I and Case II, respectively. It corresponds exactly to the unit cells presented in the theory above. The parameters of electronic components are the same to the theoretical design. Details of the sample fabrication and experimental measurements are given in Methods.

In Fig. 5(d) and 5(e), we plot the measured voltage distributions at the resonance frequency for Case I and Case II, respectively. The dominant voltage signals on sublattices in the unit cell plaquette are similar to the simulation results shown in Fig. 4(c) and 4(d). The detailed comparisons for theoretical and experimental results are provided in S1 of Supplementary Materials. It is seen clearly that the experiment results agree well with the simulation results and the hybrid 3D STT and SSS effects have also been confirmed in circuit experiments.

In summary, we have theoretically designed and experimentally constructed 2D and 3D nonreciprocal topolectric circuits for observing hybrid higher-order topological skin effects for the first time. These hybrid states are not just trivial combinations of skin states and higher-order topological states put together, even though their separate realizations are already achievements in themselves. By direct circuit simulations and voltage measurements, the 2D ST modes and 3D STT modes have not only been demonstrated experimentally, but also the corresponding 2D SS and 3D SSS effects have been found. The qualitative distinction across these modes have been clearly observed from the distribution of voltages in circuits. Our investigations show that the circuits can offer feasible platforms to investigate the exciting interplay of topology and nonreciprocity in high dimensions, which have important applications like topological switching and sensing in non-Hermitian systems.

**Methods**

**The Hamiltonian for the lattice model.**

*Hamiltonian for 2D lattice model.* In order to present the correspondence between the electric design in the main text and the lattice model, we provide the lattice Hamiltonian. The details of unit cell are shown in S2(A) of Supplementary Materials. The 2D lattice Hamiltonian can be expressed as

$$\begin{aligned} H_{2D} = \sum_{x,y} & (t_x - \delta_1) a_{x,y}^\dagger c_{x,y} + (t_x + \delta_1) c_{x,y}^\dagger a_{x,y} + (t_x + \delta_2) d_{x,y}^\dagger b_{x,y} + (t_x - \delta_2) b_{x,y}^\dagger d_{x,y} \\ & + (t_y + \delta_3) a_{x,y}^\dagger d_{x,y} + (t_y - \delta_3) d_{x,y}^\dagger a_{x,y} + (-t_y - \delta_4) b_{x,y}^\dagger c_{x,y} + (-t_y + \delta_4) c_{x,y}^\dagger b_{x,y} \\ & + t'(a_{x+1,y}^\dagger c_{x,y} + a_{x+1,y} c_{x,y}^\dagger + d_{x+1,y}^\dagger b_{x,y} + d_{x+1,y} b_{x,y}^\dagger) \\ & + t'(a_{x,y+1}^\dagger d_{x,y} + a_{x,y+1} d_{x,y}^\dagger - c_{x,y+1}^\dagger b_{x,y} - c_{x,y+1} b_{x,y}^\dagger) \end{aligned} \quad (4)$$

Here, the annihilated operator $\Lambda_{x,y} (\Lambda = a,b,c,d)$ denotes the elimination of one excitation at the sublattice $\Lambda$ of the position $(x,y)$ in the lattice, and the generated operator $\Lambda_{x,y}^\dagger (\Lambda = a,b,c,d)$ describes that one excitation at the sublattice $\Lambda$ of the position $(x,y)$. The coupling strengths between different sublattices are governed by $t_x$, $t_y$, $t'$ and $\delta_{i=1,2,3,4}$. The nonreciprocity in this 2D system depends on the value of $\delta_{i=1,2,3,4}$. We can derive the lattice Hamiltonian in the moment space

$$\hat{H}_{2D} = \sum_k \hat{\eta}_k^\dagger \begin{pmatrix} 0 & 0 & t_x - \delta_1 + t'e^{-ik_x} & t_y + \delta_3 + t'e^{-ik_y} \\ 0 & 0 & -t_y - \delta_4 - t'e^{ik_y} & t_x - \delta_2 + t'e^{ik_x} \\ t_x + \delta_1 + t'e^{ik_x} & -t_y + \delta_4 - t'e^{-ik_y} & 0 & 0 \\ t_y - \delta_3 + t'e^{ik_y} & t_x + \delta_2 + t'e^{-ik_x} & 0 & 0 \end{pmatrix} \hat{\eta}_k \quad (5)$$

where $\hat{\eta}_k^\dagger = \left( \hat{a}_k^\dagger, \hat{b}_k^\dagger, \hat{c}_k^\dagger, \hat{d}_k^\dagger \right)^T$. When nonreciprocities cancel in both directions $(\delta_1 = \delta_2, \delta_3 = \delta_4)$, no skin effect is observed in either direction. But, skin modes are still observed when OBCs are taken in both directions. We thus obtain hybrid skin-topological modes. Such lattice Hamiltonian shows the first-order topological modes when we take open and periodic boundary condition at x and y directions, respectively. Each topological mode mainly locates at two of four sublattices along the boundary, and the local nonreciprocity among these two sublattices can affect the distributions of boundary modes when both directions are open. When nonreciprocities along both directions of each plaquette do not destructively interfere $(\delta_1 \neq \delta_2, \delta_3 \neq \delta_4)$, we obtain SS modes.

*Hamiltonian for 3D lattice model*. In our description of Fig. 4(a) and 4(b), the 3D electrical architectures are stacked by 2D electric boards. These 3D electric circuits have the one-to-one correspondence to the lattices models. Here, we provide how to realize these 3D lattices by stacking 2D layers of lattices. The details of unit cell are shown in S2(B) of Supplementary Materials. The Hamiltonians for each 2D layers can be attributed to Eq. (4) above. Since different 2D layers are connected, the total Hamiltonian for such 3D lattice is

$$H_{3D} = \sum_{x,y,z} H_{2D}^z$$
$$+ (t_a + \delta_a) a_{x,y,z+1}^\dagger a_{x,y,z} + (t_a - \delta_a) a_{x,y,z+1} a_{x,y,z}^\dagger + (t_b + \delta_b) b_{x,y,z+1}^\dagger b_{x,y,z} + (t_b - \delta_b) b_{x,y,z+1} b_{x,y,z}^\dagger$$
$$+ (t_c + \delta_c) c_{x,y,z+1}^\dagger c_{x,y,z} + (t_c - \delta_c) c_{x,y,z+1} c_{x,y,z}^\dagger + (t_d + \delta_d) d_{x,y,z+1}^\dagger d_{x,y,z} + (t_d - \delta_d) d_{x,y,z+1} d_{x,y,z}^\dagger$$

(6)

Here, the annihilated operator $\Lambda_{x,y,z} (\Lambda = a,b,c,d)$ denotes the elimination of one excitation at the sublattice $\Lambda$ of the position $(x,y,z)$ in the lattice, and the generated operator $\Lambda_{x,y,z}^\dagger (\Lambda = a,b,c,d)$ describes that one excitation at the sublattice $\Lambda$ of the position $(x,y,z)$. We can derive the lattice Hamiltonian in the moment space. The lattice Hamiltonian can be expressed as

$$\hat{H}_{3D} = \sum_k \hat{\eta}_k^\dagger \begin{pmatrix} (t_a - \delta_a)e^{iq_z} + (t_a + \delta_a)e^{-iq_z} & 0 & t_x - \delta_1 + t'e^{-ik_x} & t_y + \delta_3 + t'e^{-ik_y} \\ 0 & (t_b - \delta_b)e^{iq_z} + (t_b + \delta_b)e^{-iq_z} & -t_y - \delta_4 - t'e^{ik_y} & t_x - \delta_2 + t'e^{ik_x} \\ t_x + \delta_1 + t'e^{ik_x} & -t_y + \delta_4 - t'e^{-ik_y} & (t_c - \delta_c)e^{iq_z} + (t_c + \delta_c)e^{-iq_z} & 0 \\ t_y - \delta_3 + t'e^{ik_y} & t_x + \delta_2 + t'e^{-ik_x} & 0 & (t_d - \delta_d)e^{iq_z} + (t_d + \delta_d)e^{-iq_z} \end{pmatrix} \hat{\eta}_k$$

(7)

The sublattices at the neighboring layers are connected by the coupling strengths $t_j \pm \delta_j (j = a,b,c,d)$ where $\delta_{j=a,b,c,d}$ indicates the nonreciprocity along z-direction of system.

**Theoretical model of the designed electric circuits.**

*Circuit Laplacian for 2D electric circuit.* Here, we focus on the correspondence between the non-Hermitian lattice model and our designed electric circuits. According to the Kirchhoff's law, the relation between the node alternating current and voltage should satisfy the following equation:

$$I_\alpha = i\omega^{-1}[-\sum_\beta \frac{(V_\alpha - V_\beta)}{L_{\alpha\beta}} + \omega^2 V_\alpha C_\alpha + \sum_\gamma C_{\alpha\gamma} \omega^2 (V_\alpha - V_\gamma)] \tag{8}$$

where $I_\alpha$ and $V_\alpha$ are the net current and voltage of node $\alpha$ with angular frequency being $\omega$. $L_{\alpha\beta}$ is the inductance between node $\alpha$ and node $\beta$. $C_\alpha$ is the ground capacitance at node $\alpha$. $C_{\alpha\gamma}$ is the capacitance between node $\alpha$ and node $\gamma$. The summation is taken over all nodes, which are connected to node $\alpha$ through an inductor or a capacitor.

As we can see in Fig.1(a)-(c), each unit cell possesses four nodes. In this case, the voltage and current at the site $\alpha$ should be registered as $V_\alpha (\alpha = a,b,c,d)$ and $I_\alpha (\alpha = a,b,c,d)$. Additionally, each site is connected with other sites through three kinds of coupling: intra-cell couplings ($C_1$ or $L_1$), inter-cell couplings (C or L), non-Hermitian couplings (INIC). Also, each site is grounded through capacitances and inductances to guarantee the resonance frequency. Details of the grounding part is provided in the S4 of the Supplementary Materials. Here, based on the Kirchhoff's law, we give the details of realizing the electric design to observe the SS mode in Fig.1(c). Consider the model shown in S2(C) of Supplementary Materials, there are voltages $V_a - V_d$ at the four nodes and currents $I_1 - I_8$ at the circuit branches. Using the Kirchhoff's current formula, we have

$$\begin{aligned} I_3 + I_8 &= I_6 e^{-iq_x} + I_1 e^{-iq_y} + G_a \\ I_2 + I_5 &= I_4 + I_7 + G_b \\ I_4 + I_6 &= I_8 + I_2 e^{-iq_y} + G_c \\ I_1 + I_7 &= I_3 + I_5 e^{-iq_x} + G_d \end{aligned} \tag{9}$$

where $G_i (i=a,b,c,d)$ is grounding part, $q_x$ and $q_y$ denote the phase of Block wave vector propagating in the x and y directions, respectively. So we can write the currents that flow into each node as,

$$I_a = i\omega^{-1}[-\omega^2 V_a^{x,y}\left(C_3 - \frac{2}{\omega^2 L_1} - \frac{1}{\omega^2 L_2} - \frac{2}{\omega^2 L}\right) + \omega^2(C_1 - C_3)(V_c^{x,y} - V_a^{x,y}) + \omega^2(C_1 + C_2)(V_d^{x,y} - V_a^{x,y})$$
$$+ \omega^2 C(V_d^{x,y+1} - V_a^{x,y}) + \omega^2 C(V_c^{x+1,y} - V_a^{x,y})]$$

$$I_b = i\omega^{-1}[-\frac{(V_c^{x,y} - V_b^{x,y})}{L_1} + \frac{(V_c^{x,y} - V_b^{x,y})}{L_2} - \frac{(V_c^{x,y-1} - V_b^{x,y})}{L} - \omega^2 V_b^{x,y}\left(-\frac{1}{\omega^2 L_3} - \frac{1}{\omega^2 L_2}\right)$$
$$+ (C_1 + C_3)\omega^2(V_d^{x,y} - V_b^{x,y}) + C\omega^2(V_d^{x-1,y} - V_b^{x,y})]$$

$$I_c = i\omega^{-1}[-\frac{(V_b^{x,y}-V_c^{x,y})}{L_1} - \frac{(V_b^{x,y}-V_c^{x,y})}{L_2} - \frac{(V_b^{x,y+1}-V_c^{x,y})}{L} - \omega^2 V_c^{x,y}\left(-\frac{1}{\omega^2 L_3} - \frac{1}{\omega^2 L_2}\right) \quad (10)$$
$$+ (C_1+C_3)\omega^2(V_a^{x,y}-V_c^{x,y}) + C\omega^2(V_a^{x-1,y}-V_c^{x,y})]$$

$$I_d = i\omega^{-1}[-\omega^2 V_d^{x,y}\left(-\frac{2}{\omega^2 L_1}-\frac{2}{\omega^2 L}+C_2+C_3\right)$$
$$+(C_1-C_2)\omega^2(V_a^{x,y}-V_d^{x,y})+C\omega^2(V_a^{x,y-1}-V_d^{x,y})+C\omega^2(V_b^{x+1,y}-V_d^{x,y})+(C_1-C_3)\omega^2(V_b^{x,y}-V_d^{x,y})]$$

According to the circuit Laplacian $J(\omega)$, we can write Eq.(10) in the form below,

$$\begin{pmatrix} I_a \\ I_b \\ I_c \\ I_d \end{pmatrix} = J \begin{pmatrix} V_a \\ V_b \\ V_c \\ V_d \end{pmatrix} \quad (11)$$

So the circuit Laplacian matrix $J(\omega)$ for SS mode can be expressed as:

$$J_{SS2D}(\omega) = $$

$$i\omega \begin{bmatrix} \frac{2}{\omega^2 L_1}+\frac{1}{\omega^2 L_2}+\frac{2}{\omega^2 L}-2C_1-C_2-2C & 0 & C_1-C_3+Ce^{-iq_x} & C_1+C_2+Ce^{-iq_y} \\ 0 & \frac{1}{\omega^2 L_3}+\frac{1}{\omega^2 L_1}+\frac{1}{\omega^2 L}-C_1-C_3-C & -\frac{1}{\omega^2 L_1}+\frac{1}{\omega^2 L_2}-\frac{e^{iq_y}}{\omega^2 L} & C_1+C_3+Ce^{iq_x} \\ C_1+C_3+Ce^{iq_x} & -\frac{1}{\omega^2 L_1}-\frac{1}{\omega^2 L_2}-\frac{e^{-iq_y}}{\omega^2 L} & \frac{1}{\omega^2 L_3}+\frac{1}{\omega^2 L_1}+\frac{1}{\omega^2 L}-C_1-C_3-C & 0 \\ C_1-C_2+Ce^{iq_y} & C_1-C_3+Ce^{-iq_x} & 0 & \frac{2}{\omega^2 L_1}+\frac{2}{\omega^2 L}-2C_1-2C \end{bmatrix}$$

(12)

Besides, the circuit Laplacian of hybrid ST mode can also be derived in this way. When $\omega = \omega_0 = (LC)^{-1/2} = (L_1C_1)^{-1/2} = (L_2C_2)^{-1/2} = (L_3C_3)^{-1/2}$. The circuit Laplacian matrix $J_{2D}(\omega)$ can be expressed as

$$J_{2D}(\omega) = i\omega \begin{bmatrix} 0 & 0 & C_1-C_3+Ce^{-iq_x} & C_1\pm C_2+Ce^{-iq_y} \\ 0 & 0 & -C_1+C_2-Ce^{iq_y} & C_1\pm C_3+Ce^{iq_x} \\ C_1+C_3+Ce^{iq_x} & -C_1-C_2-Ce^{-iq_y} & 0 & 0 \\ C_1\mp C_2+Ce^{iq_y} & C_1\mp C_3+Ce^{-iq_x} & 0 & 0 \end{bmatrix} \quad (13)$$

If we choose − in $\pm$ and + in $\mp$, Eq.(13) represents the circuit Laplacian for the case where the hybrid second-order ST effect can be shown. Otherwise, it corresponds to the case that can show the SS effect. Eq.(13) is Eq. (2) in the main text of the paper. It is similar to theoretical design shown in Eq.(5). Based on the consistence for the mathematical formula, it is straightforward to infer that we can implement the hybrid second-order ST effect and SS effect by using our designed electric circuits.

*Circuit Laplacian for 3D electric circuit.* As for our 3D circuit in Fig.4, each unit cell also possesses four nodes (a, b, c and d). We need to consider the coupling in z direction. Here, based on the Kirchhoff's law, we provide the details of electric design to observe the SSS mode. In this case, the Kirchhoff equation on each site can be expressed as:

$$I_a = i\omega^{-1}[-\omega^2 V_a^{x,y,z}\left(C_3 - \frac{4}{\omega^2 L_1} - \frac{1}{\omega^2 L_2} - \frac{2}{\omega^2 L}\right) + \omega^2(C_1 - C_3)(V_c^{x,y,z} - V_a^{x,y,z})$$
$$+ \omega^2(C_1 + C_2)(V_d^{x,y,z} - V_a^{x,y,z}) + \omega^2 C(V_d^{x,y+1,z} - V_a^{x,y,z}) + \omega^2 C(V_c^{x+1,y,z} - V_a^{x,y,z})$$
$$+ \omega^2(C_1 - C_2)(V_a^{x,y,z-1} - V_a^{x,y,z}) + \omega^2(C_1 + C_2)(V_a^{x,y,z+1} - V_a^{x,y,z})]$$

$$I_b = i\omega^{-1}[-\frac{(V_c^{x,y,z} - V_b^{x,y,z})}{L_1} - \frac{(V_c^{x,y,z} - V_b^{x,y,z})}{L_2} - \frac{(V_c^{x,y-z} - V_b^{x,y,z})}{L}$$
$$- \omega^2 V_b^{x,y,z}\left(-\frac{2}{\omega^2 L_1} - \frac{1}{\omega^2 L_3} - \frac{1}{\omega^2 L_2}\right) + (C_1 + C_3)\omega^2(V_d^{x,y,z} - V_b^{x,y,z})$$
$$+ C\omega^2(V_d^{x-1,y,z} - V_b^{x,y,z}) + \omega^2(C_1 - C_2)(V_b^{x,y,z-1} - V_b^{x,y,z}) + \omega^2(C_1 + C_2)(V_b^{x,y,z+1} - V_b^{x,y,z})]$$

$$I_c = i\omega^{-1}[-\frac{(V_b^{x,y,z} - V_c^{x,y,z})}{L_1} + \frac{(V_b^{x,y,z} - V_c^{x,y,z})}{L_2} - \frac{(V_b^{x,y-1,z} - V_c^{x,y,z})}{L}$$
$$- \omega^2 V_c^{x,y,z}\left(-\frac{2}{\omega^2 L_1} - \frac{1}{\omega^2 L_3} - \frac{1}{\omega^2 L_2}\right) + (C_1 + C_3)\omega^2(V_a^{x,y,z} - V_c^{x,y,z})$$
$$+ C\omega^2(V_a^{x-1,y,z} - V_c^{x,y,z}) + \omega^2(C_1 - C_2)(V_c^{x,y,z-1} - V_c^{x,y,z}) + \omega^2(C_1 - C_2)(V_c^{x,y,z-1} - V_c^{x,y,z})]$$

$$I_d = i\omega^{-1}[-\omega^2 V_d^{x,y,z}\left(-\frac{4}{\omega^2 L_1} - \frac{2}{\omega^2 L} + C_2 + C_3\right) + (C_1 - C_2)\omega^2(V_a^{x,y,z} - V_d^{x,y,z})$$
$$+ C\omega^2(V_a^{x,y-1,z} - V_d^{x,y,z}) + C\omega^2(V_b^{x+1,y,z} - V_d^{x,y,z}) + (C_1 - C_3)\omega^2(V_b^{x,y,z} - V_d^{x,y,z})$$
$$+ \omega^2(C_1 + C_2)(V_d^{x,y,z-1} - V_d^{x,y,z}) + \omega^2(C_1 + C_2)(V_d^{x,y,z-1} - V_d^{x,y,z})]$$

(14)

So the circuit Laplacian matrix $J(\omega)$ for SSS mode can be expressed as:

$$J_{SSS3D}(\omega) = i\omega \begin{bmatrix} A_1 & 0 & C_1 - C_3 + Ce^{-iq_x} & C_1 + C_2 + Ce^{-iq_y} \\ 0 & A_2 & -\frac{1}{\omega^2 L_1} - \frac{1}{\omega^2 L_2} - \frac{e^{iq_y}}{\omega^2 L} & C_1 + C_3 + Ce^{iq_x} \\ C_1 + C_3 + Ce^{iq_x} & -\frac{1}{\omega^2 L_1} + \frac{1}{\omega^2 L_2} - \frac{e^{-iq_y}}{\omega^2 L} & A_3 & 0 \\ C_1 - C_2 + Ce^{iq_y} & C_1 - C_3 + Ce^{-iq_x} & 0 & A_4 \end{bmatrix} \quad (15)$$

The diagonal elements $A_1 = \frac{4}{\omega^2 L_1} + \frac{1}{\omega^2 L_2} + \frac{2}{\omega^2 L} - 4C_1 - C_2 - 2C + (C_1 - C_2)e^{iq_z} + (C_1 + C_2)e^{-iq_z}$,

$A_2 = A_3 = \frac{1}{\omega^2 L_3} + \frac{3}{\omega^2 L_1} + \frac{1}{\omega^2 L} - 3C_1 - C_3 - C + (C_1 - C_2)e^{iq_z} + (C_1 + C_2)e^{-iq_z}$ and

$A_4 = \frac{4}{\omega^2 L_1} + \frac{2}{\omega^2 L} - 4C_1 - 2C + (C_1 - C_2)e^{iq_z} + (C_1 + C_2)e^{-iq_z}$. When the electrical components satisfy $\omega = \omega_0 = (LC)^{-1/2} = (L_1 C_1)^{-1/2} = (L_2 C_2)^{-1/2} = (L_3 C_3)^{-1/2}$, Eq.(15) becomes,

$J_{SSS3D}(\omega) =$

$i\omega \begin{bmatrix} (C_1 - C_2)e^{iq_z} + (C_1 + C_2)e^{-iq_z} & 0 & C_1 - C_3 + Ce^{-iq_x} & C_1 + C_2 + Ce^{-iq_y} \\ 0 & (C_1 - C_2)e^{iq_z} + (C_1 + C_2)e^{-iq_z} & -C_1 - C_2 - Ce^{iq_y} & C_1 + C_3 + Ce^{iq_x} \\ C_1 + C_3 + Ce^{iq_x} & -C_1 + C_2 - Ce^{-iq_y} & (C_1 - C_2)e^{iq_z} + (C_1 + C_2)e^{-iq_z} & 0 \\ C_1 - C_2 + Ce^{iq_y} & C_1 - C_3 + Ce^{-iq_x} & 0 & (C_1 - C_2)e^{iq_z} + (C_1 + C_2)e^{-iq_z} \end{bmatrix}$

(16)

The diagonal elements in the matrix (Eq. 16) represent the coupling along z direction, and the off-diagonal elements in the matrix denote the coupling on the x-y surface. It is clearly visible that the nonreciprocities appear along x, y and z directions. Following the similar construction, we can get $J_{STT3D}(\omega)$ as

$J_{STT3D}(\omega) =$

$i\omega \begin{bmatrix} (C_1 - C_2)e^{iq_z} + (C_1 + C_2)e^{-iq_z} & 0 & C_1 + Ce^{-iq_x} & C_1 + Ce^{-iq_y} \\ 0 & (C_1 - C_2)e^{iq_z} + (C_1 + C_2)e^{-iq_z} & -C_1 - Ce^{iq_y} & C_1 + Ce^{iq_x} \\ C_1 + Ce^{iq_x} & -C_1 - Ce^{-iq_y} & (C_1 + C_2)e^{iq_z} + (C_1 - C_2)e^{-iq_z} & 0 \\ C_1 + Ce^{iq_y} & C_1 + Ce^{-iq_x} & 0 & (C_1 + C_2)e^{iq_z} + (C_1 - C_2)e^{-iq_z} \end{bmatrix}$

(17)

The circuit Laplacian matrix $J_{3D}(\omega)$ can be expressed as

$J_{STT3D(SSS3D)}(\omega) = J_{TT2D(SS2D)}(\omega)$

$+ i\omega \begin{bmatrix} (C_1 - C_2)e^{iq_z} + (C_1 + C_2)e^{-iq_z} & 0 & 0 & 0 \\ 0 & (C_1 - C_2)e^{iq_z} + (C_1 + C_2)e^{-iq_z} & 0 & 0 \\ 0 & 0 & (C_1 \pm C_2)e^{iq_z} + (C_1 \mp C_2)e^{-iq_z} & 0 \\ 0 & 0 & 0 & (C_1 \pm C_2)e^{iq_z} + (C_1 \mp C_2)e^{-iq_z} \end{bmatrix}$

(18)

with

$$J_{TT2D}(\omega) = i\omega \begin{bmatrix} 0 & 0 & C_1 + Ce^{-iq_x} & C_1 + Ce^{-iq_y} \\ 0 & 0 & -C_1 - Ce^{iq_y} & C_1 + Ce^{iq_x} \\ C_1 + Ce^{iq_x} & -C_1 - Ce^{-iq_y} & 0 & 0 \\ C_1 + Ce^{iq_y} & C_1 + Ce^{-iq_x} & 0 & 0 \end{bmatrix}. \quad (19)$$

If we choose + in $\pm$ and - in $\mp$, Eq.(18) represents the circuit Laplacian for Case I. Otherwise, it corresponds to Case II. It is Eq. (3) in the main text of the paper. It is similar to theoretical design shown in Eq.(7). As revealed in this matrix (Eq. (18)), for $J_{STT3D}(\omega)$, the nonreciprocity only emerges in the diagonal elements containing $q_z$. It will bring about the corner modes driven unbalanced along z direction (this is shown in Fig. 4 and Fig. 5 of the main text). Based on the consistence for the mathematical formula, it is straightforward to infer that we can implement the hybrid 3D STT effect and 3D SSS effect by using our designed electric circuits.

**Sample fabrications and circuit measurements.** We exploit the electric circuits by using PADs program software, where the PCB composition, stackup layout, internal layer and grounding design are suitably engineered. Here, each well-designed PCB possesses totally eight layers to arrange the complex conductor. It is worthy to note that the ground layer should be placed in the gap between any two layers to avoid their coupling. Moreover, all PCB traces have a relatively large width (0.5mm) to reduce the parasitic inductance and the spacing between electronic devices is also large enough (1.0mm) to avert spurious inductive coupling. On the other hand, due to the size limit of the PCB fabrication in 3D sample, we cut the whole sample into eleven pieces. Six pieces are used to display the couplings along x and y directions and the remaining five pieces for the z direction. To ensure the same grounding condition, we link the copper pillar of each sub-PCB together. As for the circuit excitation, we use NI PXle-8840 Quad-Core Embedded Controller to input 2.77MHz alternating current. SMP connectors are welded on the PCB for the signal input and circuit measurement.

Photo to show connections in the z direction is shown in S3 of Supplementary Materials. Four sublattices (a, b, c, d) represent four INICs which connect corresponding points between two x-y layers. DB9 connectors are used to connect the INICs between each two x-y layers. Lots of external wires are used for DB9 connectors. So we need to use wires with low impedance. Besides, we need to disorganize the wires to avoid parallel interference between wires. What's more, the tatal 3D circuit is a large sample, we need to use enough voltage inputs to guarantee the work of INICs.

**Data availability.** Any related experimental background information not mentioned in the text and other findings of this study are available from the corresponding author upon reasonable request.

**Acknowledgements**

This work was supported by the National key R & D Program of China under Grant No. 2017YFA0303800, the National Natural Science Foundation of China (91850205) and C. H. Lee acknowledges support from Singapore MOE Tier I grant WBS: R-144-000-435-133.

**Author contributions**

D.Y. Z. and T. C. provided the theory with the help of C.H. L., D.Y. Z. performed the experiments with the help of W.J. H., J. C. B. and H. J. S., X.D. Z. initiated and designed this research project.

**Competing interests**

The authors declare no competing interests.

**SUPPLEMENTARY MATERIALS**

**S1. The detailed comparisons for theoretical and experimental results**

In this part, we provided detailed comparisons for theoretical and experimental results in both 2D and 3D circuits. The 2D results are shown in Fig.S1. We can see that in Fig.S1(a), voltages along right and left edges in both theoretical and experimental results are topological SSH localization. By contrast, the voltages in Fig.S1(b) are skin localization. Besides, theoretical and experimental results in ST and SS effect toward the same trend.

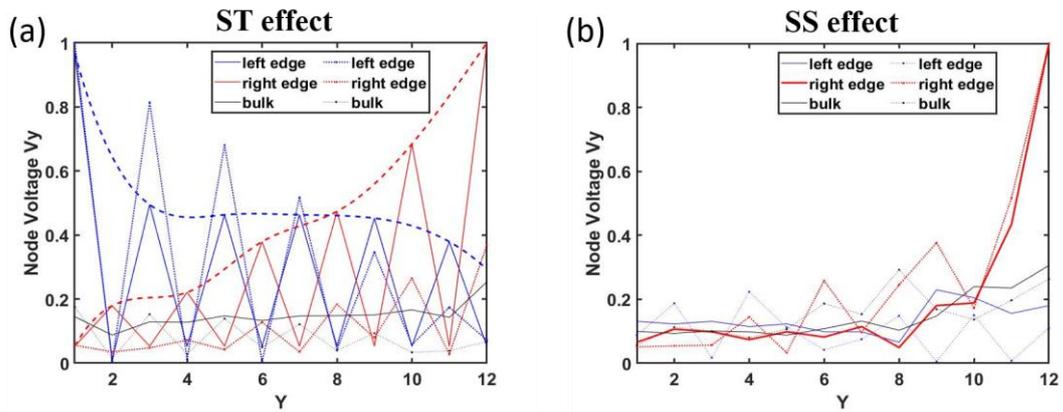

**Fig. S1.** The direct comparison for 2D theoretical and experimental results. Solid lines represent experimental results and dotted line represent theoretical results. (a) Comparison for ST theoretical and experimental results. (b) Comparison for SS theoretical and experimental results. Blue, red and black solid lines represent experimental voltage change along left edge, right edge, and average of bulk voltages in Fig.3(c) and (d), respectively. Blue, red and black dotted lines represent theoretical voltage change along left edge, right edge, and average of bulk voltages in Fig.2(c) and (d), respectively.

The 3D results are shown in Fig.S2. We can see that in Fig.S2(a), voltages in both theoretical and experimental results are topological SSH localization. By contrast, the voltages in Fig.S2(b) are skin localization. Besides, theoretical and experimental results in STT and SSS effect toward the same trend. The results mean that the experimental phenomena correspond exactly to the theoretical results, which indicates that the modes in 2D and 3D have been observed successfully in designed circuit systems.

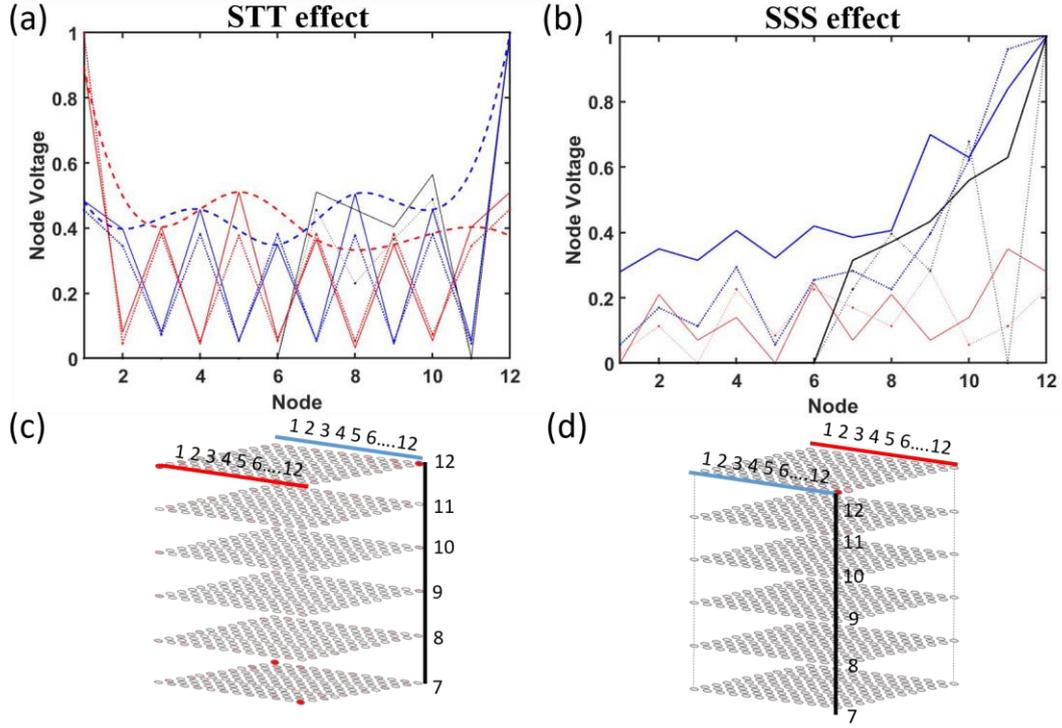

**Fig. S2.** The directly comparison for 3D theoretical and experimental results. Solid lines represent experimental results and dotted line represent theoretical results. (a) Comparison for STT theoretical and experimental results. (b) Comparison for SSS theoretical and experimental results. (c)-(d) show the corresponding colors for solid lines in (a) and (b), respectively.

## S2. Figures for the method
### A. Figure for 2D lattice model

In order to present the correspondence between the electric design in the main text and the lattice model, we provide the lattice Hamiltonian in Fig. S3. The green, blue, yellow and red spheres represent the sublattice a, b, c and d within unit cell plaquette. The details of unit cell are shown in Fig.S3.

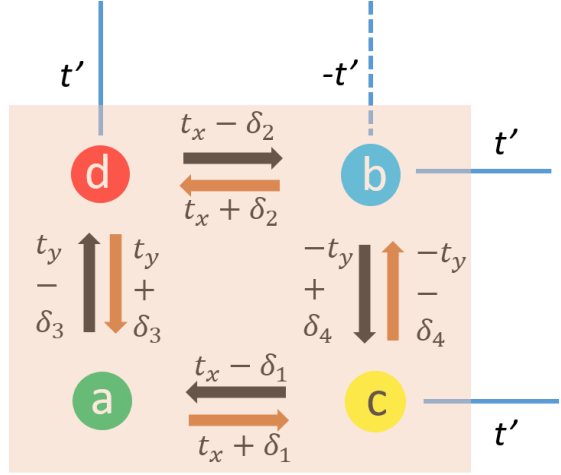

**Fig. S3.** The unit cell for the 2D lattice model. Each unit cell contains four sublattices (a, b, c, d). The brown and black lines with arrows represent asymmetric intra-cell coupling and the blue lines denote inter-cell coupling. This tight-binding model can demonstrate both hybrid second-order ST effect and SS effect with different values of $\delta_i (i=1,2,3,4)$.

## B. Figure for 3D lattice model

These 3D electric circuits have the one-to-one correspondence to the lattices models. Here, we provide how to realize these 3D lattices by stacking 2D layers of lattices. The details of unit cell are shown in Fig.S4.

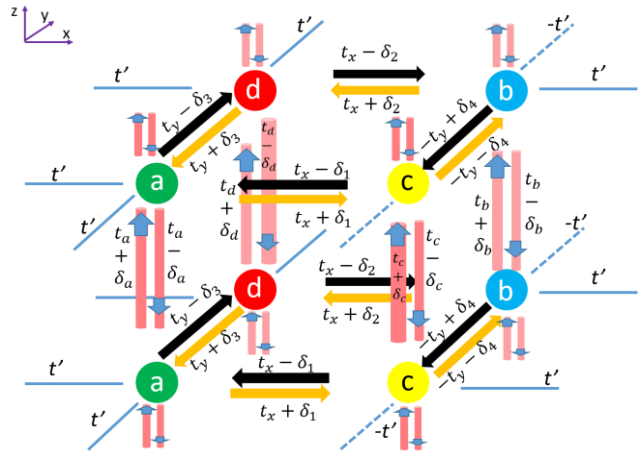

**Fig. S4.** The diagram for the 3D lattice model. We can see that each unit cell contains four sublattices (a, b, c, d). The brown and black lines with arrows represent asymmetric intra-cell coupling and the blue lines denote inter-cell coupling. The red lines with arrows represent asymmetric z direction coupling. This tight-binding model can demonstrate both hybrid 3D STT effect and 3D SSS effect with different value of $\delta_i (i=1,2,3,4,a,b,c,d)$.

## C. Figure for deriving circuit Laplacian in 2D electric circuit

Fig.S5 shows the model of unit cell for realizing the electric design to observe the SS mode in Fig.1(c). There are voltages $V_a - V_d$ at the four nodes and currents $I_1 - I_8$ at the circuit branches.

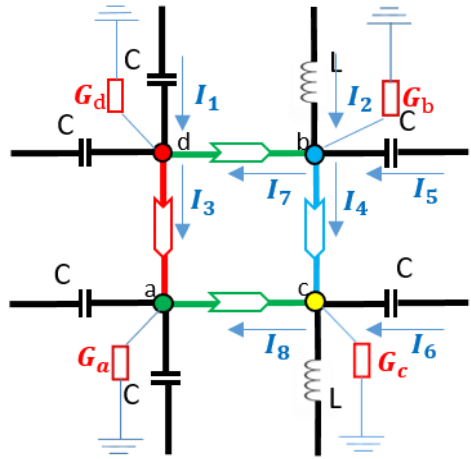

**Fig.S5.** The unit cell of proposed circuits to observe SS mode. The arrows mark the direction of currents.

## S3. Sample fabrications in the z direction

In this part, we show the details of the fabricated z-direction components for the 3D sample. The photograph are shown in Fig.S6. The right photograph shows the total INICs between two layers. The inset presents z-direction INICs in a unit cell.

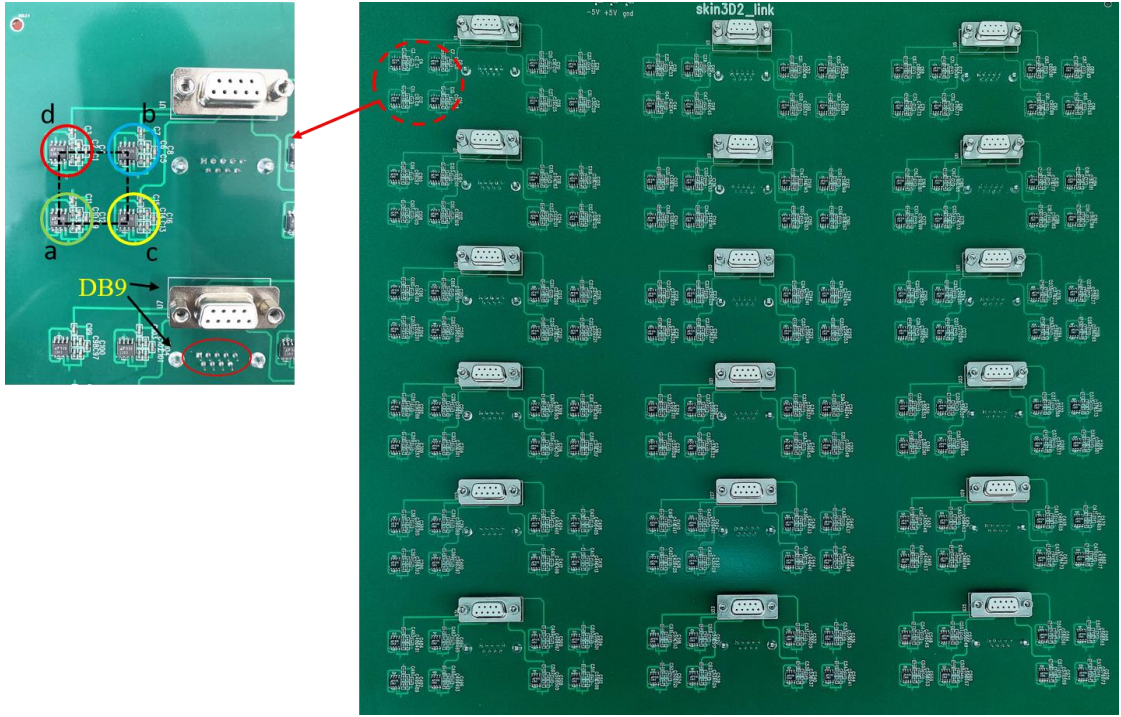

**Fig.S6.** The photograph of the fabricated z-direction components for the 3D sample. The inset presents the enlarged view. Four circles with different colors show z-direction INICs which connect corresponding points a, b, c and d between two x-y layers. We use DB9 to connect z-direction INIC with the upper and lower layers.

## S4. Grounding of the finite electric circuit

To fulfill the lattice Hamiltonian by a topolectrical circuit Laplacian, we need to choose the suitable grounding to eliminate the diagonal elements of open circuit Laplacian at the resonant frequency. This can be easily realized by making the inductivities and capacitances enter the circuit Laplacian with opposite sign, that is,

$$J(\omega) = i\omega C - \frac{i}{\omega L} \qquad (S1)$$

Hence, the contribution of inductivities (capacitances) at the fix node can be cancelled by grounding matched capacitances (inductivities). We present the grounding patterns in Tables 1-4, which correspond to the actual sample.

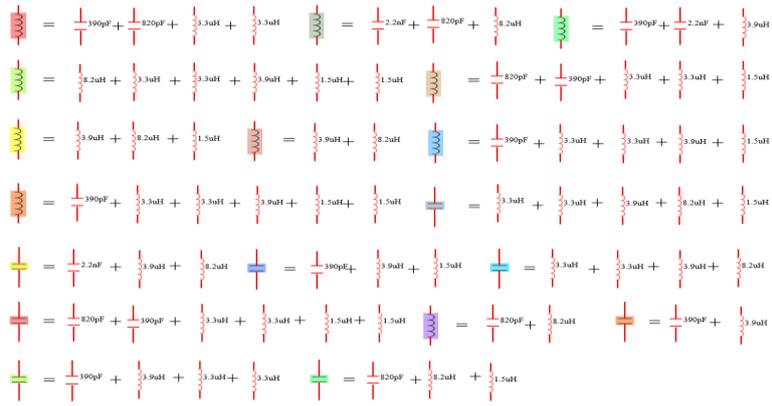

Table.1

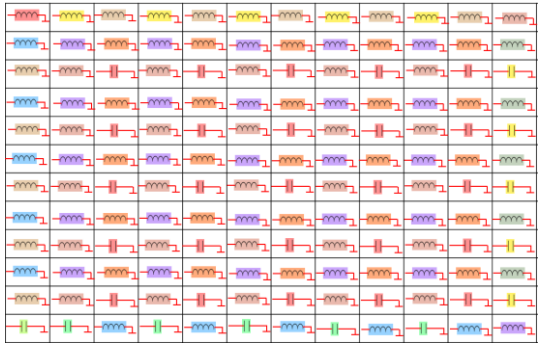

Table.2

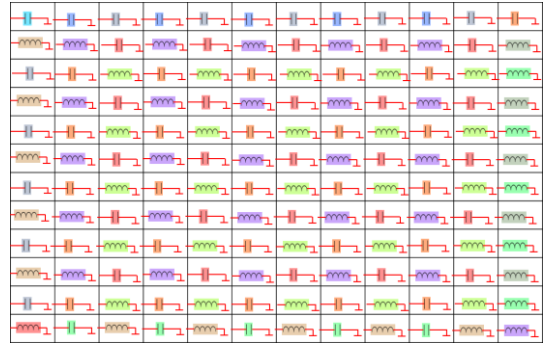

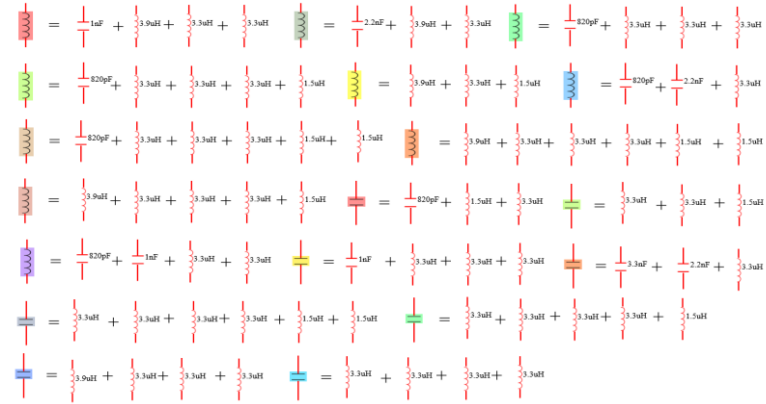

Table.3

End

Mid

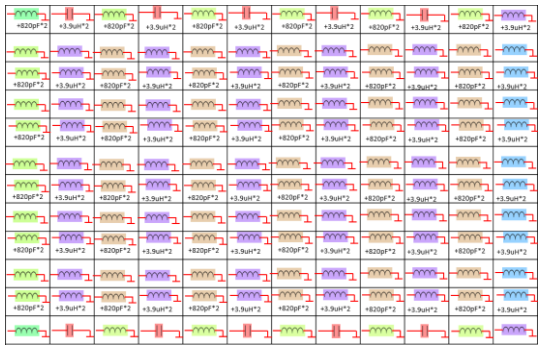

Top

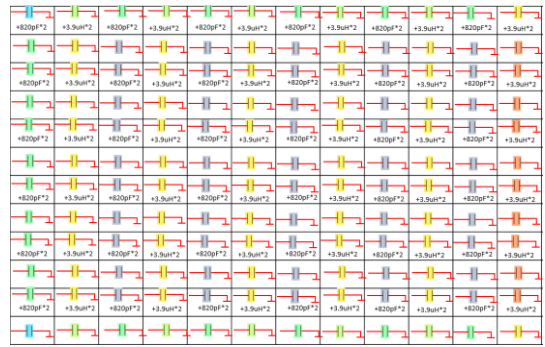

Table.4

End

Mid

Top

Table.1 is grounding part for SS effect. Table.2 is grounding part for hybrid second-order ST effect. Table.3 is grounding part for 3D SSS effect. Table.4 is grounding part for hybrid 3D STT effect. As we mentioned above, the whole 3D sample is divided to eleven layers. Only six layers have grounding parts. The end, mid and top are for the end layer, mid 2-4 layers and top layer, respectively.